\documentclass[preprint,showpacs,preprintnumbers,amsmath,amssymb]{revtex4}

\usepackage{graphicx}
\usepackage{dcolumn}
\usepackage{bm}
\usepackage{textcomp}
\newcommand {\apgt} {\ {\raise-.5ex\hbox{$\buildrel>\over\sim$}}\ }
\newcommand {\aplt} {\ {\raise-.5ex\hbox{$\buildrel<\over\sim$}}\ }
\newcommand{\addspace}{\vspace{+10mm}}

\begin{document}
 
\title{Symmetry boosts quantum computer performance}

\author{Y. S. Nam} 
\email{ynam@umd.edu}
\affiliation{Department of Computer Science, Institute for 
Advanced Computer Studies, 
and Joint Center for Quantum Information and Computer Science, 
University of Maryland, College Park, MD 20910, USA} 
\author{R. Bl\"umel}  
\affiliation{Department of Physics, Wesleyan University, 
Middletown, Connecticut 06459-0155, USA}
  
\date{\today}

\begin{abstract} 
Frequently, subroutines in quantum computers have the structure
$\mathcal{F}\mathcal{U}\mathcal{F}^{-1}$, where $\mathcal{F}$ is
some unitary transform and $\mathcal{U}$ is performing a quantum computation.
In this paper we suggest that if, in analogy to spin echoes,
$\mathcal{F}$ and $\mathcal{F}^{-1}$ can be implemented symmetrically
such that $\mathcal{F}$ and $\mathcal{F}^{-1}$ have the same hardware errors, 
a symmetry boost in the fidelity of the combined
$\mathcal{F}\mathcal{U}\mathcal{F}^{-1}$ quantum operation results. 
Running the complete gate--by--gate implemented Shor algorithm, we 
show that the fidelity boost can be as large as 
a factor 10. Corroborating and extending our numerical 
results, we present analytical scaling calculations 
that show that a symmetry boost persists 
in the practically interesting case of a large 
number of qubits. 
Our analytical calculations predict 
a minimum boost factor of about 3, valid for all qubit 
numbers, which includes the boost factor 10 observed in 
our low-qubit-number simulations. 
While we find and document this symmetry boost 
here in the case of Shor's algorithm,
we suggest that other quantum algorithms might profit from similar
symmetry-based performance 
boosts whenever $\mathcal{F}\mathcal{U}\mathcal{F}^{-1}$
sub-units of the corresponding quantum algorithm can be identified.
\end{abstract}

                       


\maketitle

\section*{Introduction}
The second half of the 20th century saw the advent
of the information technology revolution.
There is no doubt about its profound impact
on just about every aspect of modern society.
The technological innovation in computers
and networks enabled us to achieve tasks
previously thought to be impossible,
such as weather forecast, telecommunication,
the Global Positioning System, and online banking.

While the current classical technology is already impressive,
yet another revolution is about to emerge: Quantum information technology \cite{NC}.
Taking advantage of quantum superposition and entanglement,
a quantum information device is expected to be more secure and 
faster than its classical counterpart.
Epitomizing the latter is Shor's algorithm \cite{Shor,NC}, which enables us
to factor a semiprime $N = pq$, where $p$ and $q$ are
prime numbers, exponentially faster than any classical
algorithm known to date.
Shor's algorithm is often associated with code-breaking,
since semiprime factorization is at the heart
of the widely-employed Rivest-Shamir-Adleman (RSA) 
encryption algorithm \cite{RSA,NC}.

Despite all the theoretically predicted stupendous powers of quantum information devices,
we do encounter major challenges when it comes to a physical
realization of these devices: Errors and defects.
This is so, because quantum information processors are 
known to be susceptible 
to the detrimental effects of 
inexact gate operations and decoherence,
especially for a quantum computer whose workings are based on
exquisite control of quantum superposition and interference.
An early list of the potentially dangerous physical mechanisms
that may destroy the proper functioning of a quantum computer was compiled
by Landauer \cite{Landauer}, and much progress has been made to fight 
these adverse mechanisms over the past couple of decades.
For instance, overcoming the stochastic type of errors,
i.e., errors that occur on a single qubit probabilistically,
was the invention of quantum error correction \cite{Shor9,CSS1,CSS2,QECC5} 
and its fault-tolerant implementation \cite{Fault1,Fault2,Fault3},
culminating in the standard de facto approaches of
topological and surface codes (see \cite{Surface} and references therein).

Still, if we are to truly realize a working, physical quantum computer
that is practically useful, the limits of engineering must be taken into consideration.
Otherwise, a quantum computer will remain
an academically interesting device of no practical relevance.
One way to approach this problem is to investigate
the accuracy gain in logical operations on a logical qubit,
given the technology-dependent physical qubit error rate.
Pioneering work in \cite{Autotune}, for instance, 
is already making headway in this direction.
Yet, another way to help realize the full potential of quantum computing
is to investigate the algorithmic performance behavior
at the logical qubit level, providing quantum experimentalists
and engineers with a logical error-rate target,
potentially easing the physical accuracy and precision requirements.
Adding to recent experimental breakthroughs, 
such as reported in \cite{FM} and \cite{Morello},
this paper provides a powerful additional 
strategy for 
realizing a working, physical quantum computer.


\section*{Methods}
As a testbed algorithm
we chose Shor's algorithm, implemented 
according to Beauregard's architecture \cite{Beau}.
We selected this particular architecture based on the facts that
(i) Shor's algorithm is arguably the most interesting
and most important quantum algorithm to date,
(ii) the algorithm is complex enough to realistically 
capture the effects of faulty gates, and, most importantly,
and exploited in this paper,
(iii) Beauregard's architecture allows us to take
advantage of symmetry.
Whether some other Shor algorithm architectures,
such as those presented in \cite{ShorArch} (and references therein),
may be exploited in a similar fashion is currently under investigation
and the results will be reported elsewhere.

Studies addressing the effects of errors and defects
on a quantum computer running Shor's algorithm 
continue to be of central interest to many scientists.
A list of early, notable contributions includes the investigations by
Cirac and Zoller \cite{CZ} studying the effect of errors 
in interaction time and laser detuning,
Miquel {\it et al.} studying the effects of interactions with
a dissipative environment \cite{Mq1} and phase drift errors \cite{Mq2},
Wei {\it et al.} exploring the effects of coherence errors occurring
while the quantum computer is idling \cite{Wei}, and
Garc\'ia-Mata {\it et al.} simulating static imperfections in Shor's algorithm \cite{Shep}.
Recent developments in quantum simulation software \cite{Liquid, Quipu} 
reflect the fact that quantum computers remain
at the forefront of research.
Our work extends this line of research
in that we simulate the entire Shor algorithm, gate--by--gate. 
Based on this 
complete implementation of Shor's algorithm, we 
investigate 
the effects of errors in the phase-rotation gates. 

We note that our error model, to be discussed in the following, 
reflects the effects of hardware errors that are unavoidable and 
guaranteed to occur 
in any hardware that exists in nature.
This is so, because even in principle there exists no physical 
equipment that will meet 
the mathematically exact circuit specifications. 
As a consequence, even if 
the quantum computer is protected with hardware implementing 
quantum error correction circuitry
according to any 
quantum error correction protocol, each and every single 
physical quantum gate of the protection circuit will 
inevitably 
contain hardware errors.
Thus, because hardware errors affect all qubits, 
including the qubits of the correction circuitry, 
there is no type of hardware error that would be correctable. 
In fact, it can be shown (see Supplementary Material) 
that hardware errors, omnipresent everywhere in a 
quantum computer,
may be more significant than the commonly-addressed
locally stochastic errors, often thought to be the most 
significant source of instability of quantum computers. 
Our error model, therefore, includes the effects of physical errors, 
i.e. hardware errors, 
that are of prime importance for stable quantum computation and, 
as shown in the 
Supplementary Material, may indeed be more important than 
local stochastic errors. 

Since the most frequently used quantum gate in Beauregard's architecture
of Shor's algorithm is a phase rotation gate
\begin{equation}
\label{theta-ideal}
\theta_j^{(\pm)} = \left( \begin{matrix} 
				1 & 0 \\
				0 & e^{\pm i\frac{\pi}{2^j}}
			    \end{matrix} \right),
\end{equation}
which appears $\sim 18 L^4$ times throughout the algorithm \cite{NB4}
when using the minimally required number of qubits to factor a semiprime $N$
whose bit-length is $L$, we tested the sensitivity of this quantum computer
running Shor's algorithm with respect to errors 
in $\theta_j^{(\pm)}$. 
Specifically, we used a statistical error model of the rotation gate
of the form \cite{NB6} 
\begin{equation}
\label{theta-defect-R}
{}^{R} \theta_j^{(\pm)} = \left( \begin{matrix} 
				1 & 0 \\
				0 & e^{\pm i\frac{\pi}{2^j} (1+\alpha^{(\pm)})}
			    \end{matrix} \right),
\end{equation}
in the case where the errors scale 
according to the size of the gate operation and
\begin{equation}
\label{theta-defect-A}
{}^{A} \theta_j^{(\pm)} = \left( \begin{matrix} 
				1 & 0 \\
				0 & e^{\pm i\left(\frac{\pi}{2^j} + \alpha^{(\pm)}\right)}
			    \end{matrix} \right),
\end{equation}
in the case where the errors do not scale
according to the size of the operation. 
In both cases 
$\alpha^{(\pm)}$ is the defect parameter 
that may or may not be (strongly) correlated with 
the gate type indexed with $j$. In case a 
one-to-one correlation 
exists, we call the error ``typed'' and replace 
$\alpha^{(\pm)}$ with $\alpha_j^{(\pm)}$. 
The $\pm$ sign corresponds to forward and backward operation.

The reason why we explicitly 
distinguish these two error models is as follows.
First, any physical device has a finite accuracy, and
this is usually given in terms of percentage error 
with respect to the size of the gate operation.
Since a rotation gate $\theta_j$ is built
according to a gate decomposition sequence (see references in \cite{NB-qic}),
the approximated 
rotation gate will contain errors that scale in the size of the operation,
especially regarding the construction method of $\theta_j$, for instance 
by applying $\theta_{j+1}$ twice.
This iteration method may be realistic and desirable
from the technological or economical perspective.
Thus, characterizing a device in terms of relative errors is captured
by the ${}^{R}\theta_j^{(\pm)}$ model.
However, suppose we characterize our quantum computer device
in terms of its technological limit, say $\delta$.
In this case, most likely, all gates are to be made with
different sequences resulting in an error level $\lesssim \delta$, 
and this is captured by our model ${}^{A}\theta_j^{(\pm)}$.

We now subdivide both models into 3 categories:
(i) typed errors ($\alpha^{\pm} = \alpha^{\pm}_j$), 
asymmetric ($\alpha^{+} \neq \alpha^{-}$),
(ii) typed errors ($\alpha^{\pm} = \alpha^{\pm}_j$), 
symmetric ($\alpha^{+} = \alpha^{-}$), and 
(iii) non-typed errors, i.e., completely random $\alpha^{\pm}$.
The three categories are explained as follows.
Typing arises from using the same sequence, or the same physical device, 
for the same $\theta_j$ that occur multiple times throughout the entire Shor algorithm.
Then, depending on the way that the physical device is set up,
since the backward gate is nothing but a unitary inverse of the forward gate,
we may assume that the errors of the forward and backward
gates are symmetric. Therefore, 
while (i) and (ii) capture systematic errors,
(iii) deals with random errors.

\section*{Results}
\begin{figure}
\includegraphics[scale=1,angle=0]{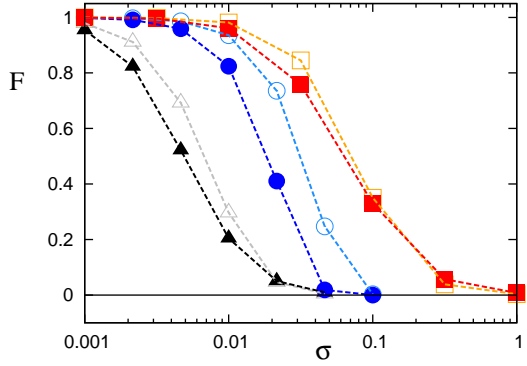}
\addspace
\caption {\label{fig1}
Fidelity $F$ of a quantum computer factoring $N=15$ with seed $2$ 
as a function of standard deviation $\sigma$ of the logical-gate errors.
The quantum computer is equipped with adders that are suitable for use
in Shor-algorithm factoring of at most 4-bit semiprimes. 
Shown are the cases of
typed, asymmetric errors (triangles),
typed, symmetric errors (squares), and
non-typed, random errors (circles).
Filled plot symbols (red, blue, and black) 
denote relative errors [see Eq.(\ref{theta-defect-R})]
and the open plot symbols (orange, cyan, and grey) 
denote absolute errors [see Eq.(\ref{theta-defect-A})].
Dashed lines connecting plot symbols are drawn to guide the eye.
The solid, horizontal line corresponds to $F = 0$. 
}
\end{figure}

To start with, we simulate the case of factoring $N=15$.
This is the case used in \cite{Mq2}, which allows us to compare 
our results with the results in \cite{Mq2}. 
Defining the fidelity 
$F = |\langle \Psi_{{\rm actual}} | \Psi_{{\rm ideal}} \rangle|^2$
as in \cite{Mq2}, in Fig.~\ref{fig1} we plot $F$ as a function of $\sigma$,
where the errors $\alpha^{(\pm)}$ are Gaussian distributed
random variables with mean 0 and standard deviation $\sigma$.
Consistent with the results presented in \cite{Mq2}, the fidelity $F$
of Shor's algorithm follows the form 
$F = \exp(-\gamma \sigma^2)$ for small $\sigma$.
At first glance, we observe that the performance of the 
quantum computer improves 
in the order of asymmetric, random, and symmetric errors.
In particular, symmetric errors give rise to a fidelity boost in $1/\gamma$
by an astonishing factor of $\sim 100$ in both the 
${}^{R}\theta_j^{(\pm)}$
and the ${}^{A}\theta_j^{(\pm)}$ models. 
In other words, to obtain comparable $F$, symmetric errors
allow for about a factor 10 larger $\sigma$.

The important question to ask now is whether the symmetry-driven fidelity boost
will persist as we scale up the quantum circuitry.
To start with, we compare the expected fidelities from naively multiplying
the fidelities of the basic building blocks of Shor's algorithm, 
i.e., the quantum adders.
This product formula of fidelities has been shown in \cite{Mq2}
to work well
in the uncorrelated random cases 
(see also the Supplementary Material).

For an $L+1$ bit sized quantum adder, capable of executing $s+a$,
where $s$ and $a$ are integers of bit length $\leq L$,
one may show that 
the phase $\Phi$ associated with $s+a$ in the symmetric case  
is given by 
\begin{equation}
\label{Phi}
\Phi_{s,a} (l) = \frac{1}{2^{L+1}} \Bigg[
1+\exp\Bigg(i\Bigg\{\Bigg[ \sum_{\nu=0}^{L-1} k_{\nu} r_{L-\nu-1}\Bigg]\Bigg\}\Bigg)
e^{2\pi i(s+a-l)/2^{L+1}} \Bigg] R_{s,a} (l),
\end{equation}
where $k_{\nu} = s_{[\nu]} + a_{[\nu]} - l_{[\nu]}$,
where $s_{[\nu]}$, e.g., denotes the $\nu$th binary digit of $s$,
\begin{equation}
\label{R}
R_{s,a}(l) = \sum_{l'=0}^{2^L-1} \exp\Bigg[i\Bigg(\sum_{m=0}^{L-1}
l'_{[L-1-m]}\Bigg\{a_{[m]}r_0 + \Bigg[ \sum_{\nu=0}^{m-1} k_{\nu}r_{m-\nu}\Bigg]
\Bigg\}\Bigg)\Bigg] e^{2\pi i (s+a-l)l'/2^L},
\end{equation}
$r_j$ may be $\alpha_j$ or $(\pi/2^j)\times \alpha_j$
if the errors are of an absolute kind or of a relative kind, respectively,
and $l$ is the output integer.
The non-typed error cases are obtained by
removing correlation via letting each term in $k_\nu$
be associated with individual random terms,
followed by removing typing of errors associated with 
the subscript $j$ of $r_j$.

Calculating now the fidelity of the quantum adder
$F_{\text{adder}} = |\Phi_{s,a}(l=s+a)|^2$,
using (\ref{Phi}) and (\ref{R}) and assuming that 
the central limit theorem holds, 
we find 
in the limit that $L$ is large and $\sigma$ is small 
\begin{align}
\label{Ratio}
&\frac{\displaystyle\min_{s,a} \Bigg|\ln\Bigg( {}^{R}F_{\text{adder}}^{\text{Typed,Sym}}\Bigg)\Bigg|}
{\langle \Bigg| \ln\Bigg( {}^{R}F_{\text{adder}}^{\text{Non-Typed}} \Bigg) \Bigg| \rangle_{s,a}}
\approx \frac{1}{3}, \qquad 
\frac{\displaystyle\max_{s,a} \Bigg|\ln\Bigg( {}^{R}F_{\text{adder}}^{\text{Typed,Sym}}\Bigg)\Bigg|}
{\langle \Bigg| \ln\Bigg( {}^{R}F_{\text{adder}}^{\text{Non-Typed}} \Bigg) \Bigg| \rangle_{s,a}}
\approx \frac{11}{18}, \\ \nonumber
&\frac{\displaystyle\min_{s,a} \Bigg|\ln\Bigg( {}^{A}F_{\text{adder}}^{\text{Typed,Sym}}\Bigg)\Bigg|}
{\langle \Bigg| \ln\Bigg( {}^{A}F_{\text{adder}}^{\text{Non-Typed}} \Bigg) \Bigg| \rangle_{s,a}}
\approx 0, \qquad 
\frac{\displaystyle\max_{s,a} \Bigg|\ln\Bigg( {}^{A}F_{\text{adder}}^{\text{Typed,Sym}}\Bigg)\Bigg|}
{\langle \Bigg| \ln\Bigg( {}^{A}F_{\text{adder}}^{\text{Non-Typed}} \Bigg) \Bigg| \rangle_{s,a}}
\approx 1.
\end{align}

We see from (\ref{Ratio}) that exploiting symmetry in our circuitry
improves the fidelity of the quantum computer.
In particular, the symmetry-driven boost always exists,
outperforming the average fidelity of the non-typed random cases
at all times asymptotically.
Based on the naive product formula of fidelities, we conclude that
the symmetry-driven fidelity boost persists in large-scale quantum circuits
that are of practical interest.

Now, the observed boost in Fig.~\ref{fig1} appears larger than 
what may be expected from (\ref{Ratio}), in particular in the 
case of relative errors.
This motivates us to find additional boost mechanisms that are not cpatured
by the naive adder-fidelity product approximation of the Shor processor fidelity.
While we were not able to pin down all boost mechanisms, we present 
in the following the one that is based on the next-level-up building blocks,
namely the modulo addition gates.

To start, we point out that a modulo-addition gate consists of
five adders and an auxiliary qubit (see, e.g., Fig.~5 of \cite{Beau}).
For an input integer value of $s$, a quantum modular addition
of $s+a \bmod N$ may be performed by first adding $a$ then
subtracting $N$, followed by a conditional operation of adding back
$N$ if $s+a < N$, which may be done with the help of an auxiliary qubit.
This completes the computational part of the modular addition.
In order to now unitarily restore and decouple the auxiliary qubit 
that is at this point in its conditional state (depending on the relation
between $s+a$ and $N$), two additional adders that
subtract and add $a$, respectively, are used.
We refer to this step as the recovery part of the modular addition.

\begin{figure}
\includegraphics[scale=0.45,angle=0]{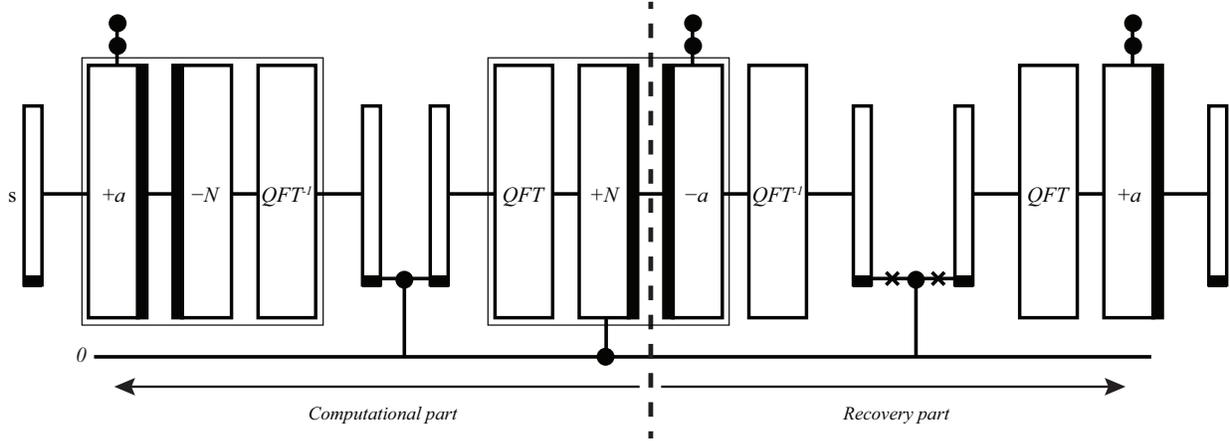}
\caption {\label{fig2}
Modulo addition gate circuit diagram, inspired by Fig.~5 of \cite{Beau}.
Circles denote controlling qubits and Xs denote NOT gates.
Thick black bars identify adders and subtractors, 
i.e., bar-right for adders 
and bar-left for subtractors. 
Black solid squares in the qubit register, denoted by thin rectangles,
denote the most significant digit qubit of the register. 
All additions and subtractions are performed in the Fourier space.
Solid grey boxes denote the symmetric parts used in the derivation
of $F_{\text{s.s.}}$ discussed in the text. The
dashed line denotes the border between
computational and recovery parts of the modulo addition circuit.
}
\end{figure}

According to whether the conditional addition of $N$ 
is triggered or not,
we consider two cases, i.e., (i) $s+a < N$ and (ii) $s+a \geq N$.
In the former case, because of the triggering,
the modulo-addition circuit attains a symmetric substructure,
denoted by the solid lines in Fig.~\ref{fig2}.
Thus, motivated by the existence of the highly organized structure 
and in the limit of small errors,
we write the fidelity of a modulo-addition gate
in case (i) as 
$F^{(\text{i})} \approx F_{\text{s.s.}} F_{\text{adder}}^{(a)}$,
where $F_{\text{s.s.}}$ denotes the fidelity associated with the symmetric
substructure and $F_{\text{adder}}^{(a)}$ denotes the fidelity of the last
adder with addend $a$ in Fig.~\ref{fig2}, all equipped with symmetric noise.
In the latter case, the auxiliary qubit is not turned on, 
resulting in the modulo addition gate fidelity of case (ii)
$F^{(\text{ii})} \approx F_{\text{adder}}^{(-N)} F_{\text{adder}}^{(a)}$,
assuming that, in the limit of small errors, the errors commute and thus
the errors associated with the first adder of the computational part
of the modulo addition gate approximately cancel those associated with
the first adder (subtractor) of the recovery part of the modulo addition gate.

At this point we notice that the only unknown term is $F_{\text{s.s.}}$,
since the fidelity of the quantum adder has already 
been discussed earlier in this paper.
Therefore, we now focus on $F_{\text{s.s.}}$.

Defining $P_{\text{remain}}$ as the probability of obtaining the ideal bit value
of the most significant qubit right after the first box in Fig.~\ref{fig2}, 
one may show $F_{\text{s.s.}} = P_{\text{remain}}^2$.
Now, $P_{\text{remain}} = \sum_{l>2^L} |\Phi_{s,a-N}(l)|^2$,
where $\Phi_{s,a-N}$ is nothing but (\ref{Phi}) with
$a_{[\nu]} \rightarrow a_{[\nu]}-N_{[\nu]}$ and $a \rightarrow a-N$.
In fact, we may write $\Phi_{s,a-N}(l)$ as 
$\cos[\pi(s+a-N-l)/2^L + \sigma^{(\nu)}] |R_{s,a-N}|/2^{L+1}$
up to a phase, where $\sigma^{(\nu)}$ is the sum in the exponent in (\ref{Phi}).
The remaining term is $|R_{s,a-N}|$, which we analyze next.

\begin{figure}
\includegraphics[scale=1,angle=0]{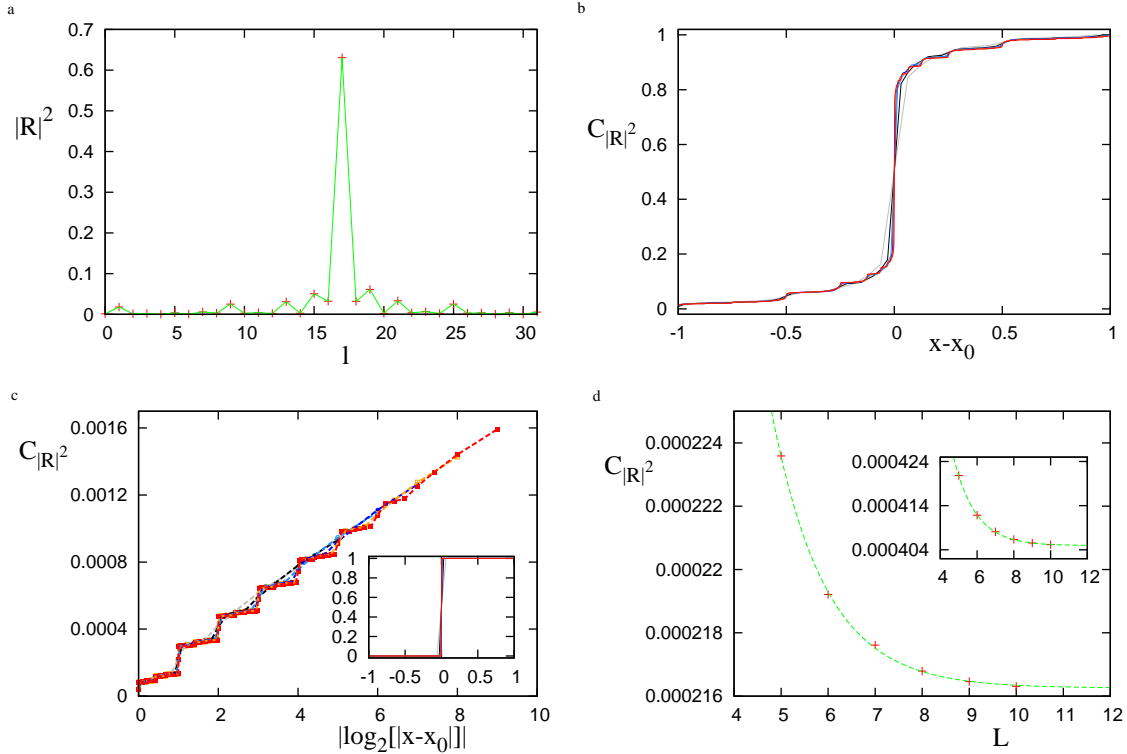}
\addspace
\caption {\label{fig3}
Various quantities related to $R$ in (\ref{R}) for relative, symmetric noise
for $s=0$, $a=0$, and $N = 2^L - 1$. a shows $|R|^2$ as a function of $l$
for $N = 2^4 - 1$, where $\sigma = 0.2$.
b shows the corresponding cumulative $|R|^2$,
$C_{|R|^2}(x = l/2^{L-1}) = [|R(l)|^2 + |R(l)|^2]/2$,
c shows $C_{|R|^2}$ for $\sigma = 0.01$
as a function of $|\log_2[|x-x_0|]|$;
inset shows an equivalent plot to b.
For both b and c, in the order of
grey, black, cyan, blue, orange, and red,
$L = 4,5, ..., 10$, respectively.
d shows an exponential convergence
of $C_{|R|^2}$ for $\log_2[|x-x_0|]| = 1$
(see inset for $\log_2[|x-x_0|]| = 2$)
as a function of $L$.
}
\end{figure}

\begin{figure}
\includegraphics[scale=1,angle=0]{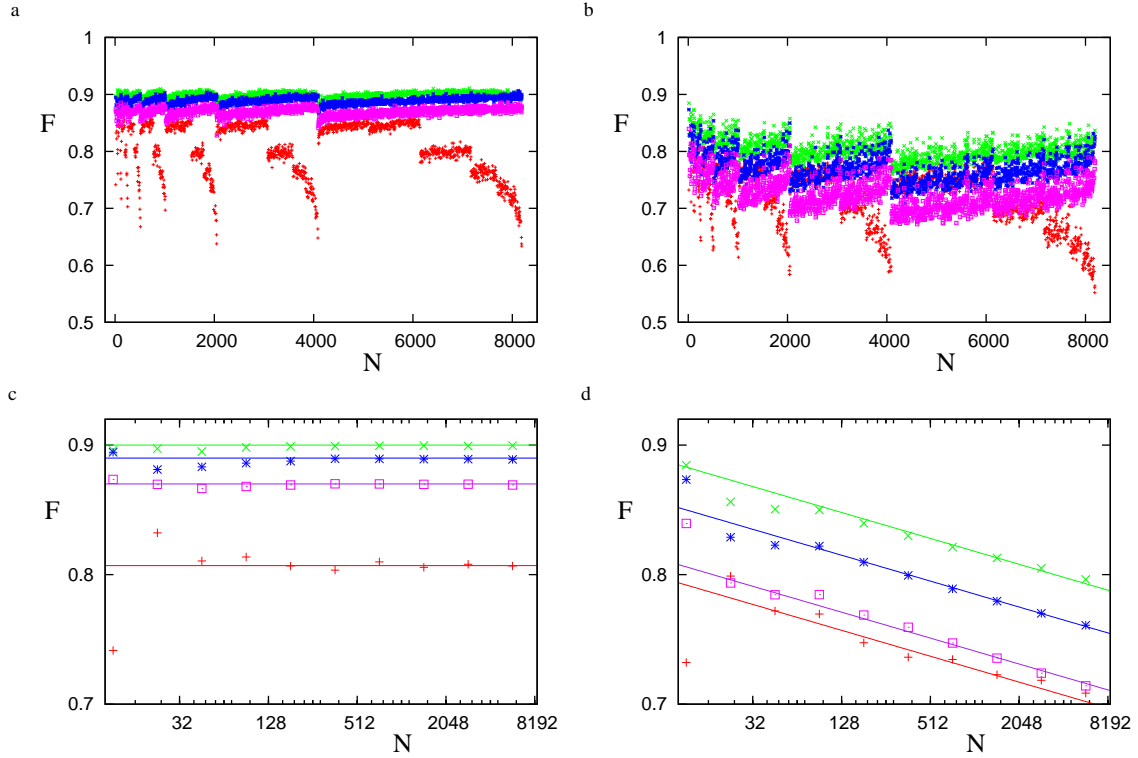}
\addspace
\caption {\label{fig4}
\small{Fidelity $F$ of a modulo-addition gate performing $0+0\bmod N$,
where $N$ is a semiprime.
Frames~a and b show $F$ as a function of
all odd semiprimes $N < 2^{13}$ for
relative and absolute symmetric errors, respectively.
The error strength used is $\sigma = 0.2$.
In the order of pluses (red), crosses (green), asterisks (blue),
and squares (purple), the adders are equipped with
0, 1, 2, and 3 additional qubits than minimally required.
Frames~c and d show logarithmically averaged $F$,
i.e., each point plotted at $2^{j+1/2}$ 
is the result of averaging over $N$
from $2^{j}$ to $2^{j+1}$, where $3 \leq j \leq 12$.
Notice that for $j=3$ and $j=4$, there is only
one semiprime each, namely, 15 and 21, respectively,
resulting in larger fluctuations due to insufficient statistics.
Solid lines in Frame~c, with corresponding color symbols,
are the tail-region fit lines $F = 0.807, 0.9, 0.89, \,\text{and}\, 0.87$.
Solid lines in Frame~d, with corresponding color symbols,
are the tail-region fit lines (to first order) $F=-0.01 \log_2 (N)+k$,
where $k=0.827, 0.918, 0.885,$ and $0.841$ for
the four cases shown.}
}
\end{figure}

In order to gain analytical insight,
we consider $s=0$, $a=0$, and $l=-N$. In this case, $R$ has
a structure where aligned phasors add up with small phase-angle
perturbations of the form $\sum_{m}-l'_{[L-2-m]}N_{[m]} \pi r_0$.
In all other cases ($l \neq -N$), the 
phasors interfere destructively with the additional
perturbation of the $\nu$-sum in (\ref{R}).
Now, because the interference without noise is perfect,
the existence of the perturbation gives rise to a non-zero, imperfect interference.
Thus, the nature of the imperfection determines $|R|$.
We find that [see Fig.~\ref{fig3} a-c for sample cases with $N=2^L-1$
and relative errors] whenever the Hamming distance 
between $-N$ and $l$ is 1 (or small), i.e., $|l-(-N)| = 2^{\mu}$, 
where $\mu$ is an integer, the magnitude $|R_{l,N}|$ is
relatively large (compared to $|l-(-N)|\neq 2^{\mu}$).
This is consistent with our analytical understanding that the more $k_\nu$'s
become non-zero, the more randomness is introduced to the perturbation angle,
resulting once again in destructive interference, but this time
of a statistical nature. 
In fact, we confirm its manifestation in the modulo addition $0+0\bmod N$
fidelity $F$ for all odd semiprimes $N<2^{13}$,
as shown in Figs.~\ref{fig4} a and b; Semiprimes $N$ between
$2^j$ and $2^{j+1}$ are sectioned into different $F$-bands,
arising from the bit-spectra of different $N$ values,
i.e., the binary digit 1 in the digit spectrum of $N$ turns on
the corresponding noisy rotation gate operation.

We also notice that, based on Fig.~\ref{fig2} c, 
$|R|$ is localized in $l$.
This is expected, since the form of $R$ in (\ref{R}) remains the same
as a function of $L$ while the associated cumulative errors are bounded
due to the exponential scaling of the error terms in $L$.
In fact, the sum of $|R|^2$ for $|l-N|<2^{L-1}$ equals 1
(see Fig.~\ref{fig2} b),
where $R(l) = R(l+2^{L+1})$.
We explicitly confirm numerically that the convergence toward the limiting,
localized distribution is exponentially better for increasing $L$ (see Fig.~\ref{fig4} d).

Together with the observed localization,
we find $P_{\rm remain}$ to be constant as a function
of increasing $L$ ($\sigma^{(\nu)}$ is bounded).
This is consistent with the plateau behavior observed in Fig.~\ref{fig3} c,
in which, to highlight the result shown in Fig.~\ref{fig3} a,
we averaged $F$ over $N$ in logarithmic scale, i.e., $2^j < N < 2^{j+1}$
for $j = 3, 4, \ldots, 12$, and plot the results
(see Fig.~\ref{fig4} d for the average results for Fig.~\ref{fig4} b).
In contrast to the relative kind of errors, the case of 
absolute errors is known to have a 
fidelity scaling that is one power less favorable in $L$ in the exponent
of fidelity (see, e.g., \cite{NB-RH}), 
and this is manifestly visible in Fig.~\ref{fig4} d.

Following the localization result demonstrated in Figs.~\ref{fig3} c and d,
assuming the fidelity $F_{\text{adder}}$ of a quantum adder
predicts the limiting distribution to a very good approximation,
we may write
\begin{equation}
\label{Rx}
|R(x=l/2^{L-1})|^2 \approx \frac{\eta}{2\ln(2)|x-x_0|}
e^{-\eta[1+\log_2(1/|x-x_0|)]},
\end{equation}
where $x_0$ is the scaled, ideal output 
and we used $F_{\text{adder}} = e^{-\eta L}$ from \cite{NB-RH}, 
where $\eta$ is a constant.
Approximating now the sum over $l>2^{L-1}$ in $P_{\text{remain}}$
as an integral, together with $|R(x)|$ in (\ref{Rx}), we obtain
\begin{equation}
\label{Premain}
P_{\text{remain}} \approx \int_0^1 \cos^2\left[ \frac{\pi(x-x_0)}{2} \right] |R(x)|^2 dx,
\end{equation}
where we assumed $\sigma^{(\nu)}$ is small.
This completes our analytical calculation for the only unknown term $F_{\text{s.s.}}$.

Equipped with our analytical 
fidelity scaling formulae, we once more 
check for the symmetry-driven fidelity boost.
For a sufficiently large quantum circuit, 
such as Shor's algorithm factoring large semiprimes
that are of practical interest, the input $s$ of a modulo addition gate
performing $s+a \bmod N$ may range anywhere between $0$ and $N-1$.
This results in an approximately 50/50 chance of (i) $s+a < N$ and (ii) $s+a \geq N$,
assuming a random $s$ and $a$ uniformly distributed between $0$ and $N-1$.
Thus, we expect the average fidelity $F_{\text{add-mod}}$ of a modulo addition gate
to be $0.5[F^{(\text{i})} + F^{(\text{ii})}]$.
Now, the addition of the addend $a$ of the modulo addition $s+a \bmod N$
occurs with probability $1/4$, assuming random bit values of the two controlling qubits
of the addition (see Fig.~5 of \cite{Beau} for detail).
Therefore, assuming once again that the product formula of fidelity holds,
this time applied to the modulo addition gate, of which there are $4L^2$ in
one complete run of Shor's algorithm, we obtain the symmetric noise Shor fidelity
\begin{equation}
\label{Fshor}
F_{\text{Shor}}^{(\text{Sym})} = F_{\text{add-mod}}^{4L^2} = 
\left( \frac{3}{4}F_{\text{s.s.}} + \frac{1}{4} 
\left[ \frac{F_{\text{s.s.}}F_{\text{adder}}+F_{\text{adder}}^2}{2} \right] \right)^{4L^2}.
\end{equation}
This may be compared to
\begin{equation}
F_{\text{Shor}}^{(\text{Non-Typed})} = 
\left[ \frac{3}{4} (F_{\text{adder}})^2 + \frac{1}{4} (F_{\text{adder}})^5\right]^{4L^2}
\end{equation}
for the uncorrelated noise counterpart. 
Importing $F_{\text{adder}}$ from Equation~(19) of \cite{NB-RH}, 
we obtain, for instance, ${}^R F_{\text{Shor}}^{(\text{Non-Typed})} = 79\%$
for $\sigma = 0.01$ and $L=4$ to leading order in $L$ in the exponent
of $F_{\text{adder}}$, in excellent agreement with Fig.~\ref{fig1}.
An equivalent computation for the symmetric case based on (\ref{Fshor}),
together with a proper normalization of (\ref{Rx}), i.e.,
$\sum_{l<2^{L-1}} |R|^2 = 1$, results in $89\%$, which is 
in satisfactory agreement with the simulation results shown in Fig.~\ref{fig1}.

\begin{figure}
\includegraphics[scale=1,angle=0]{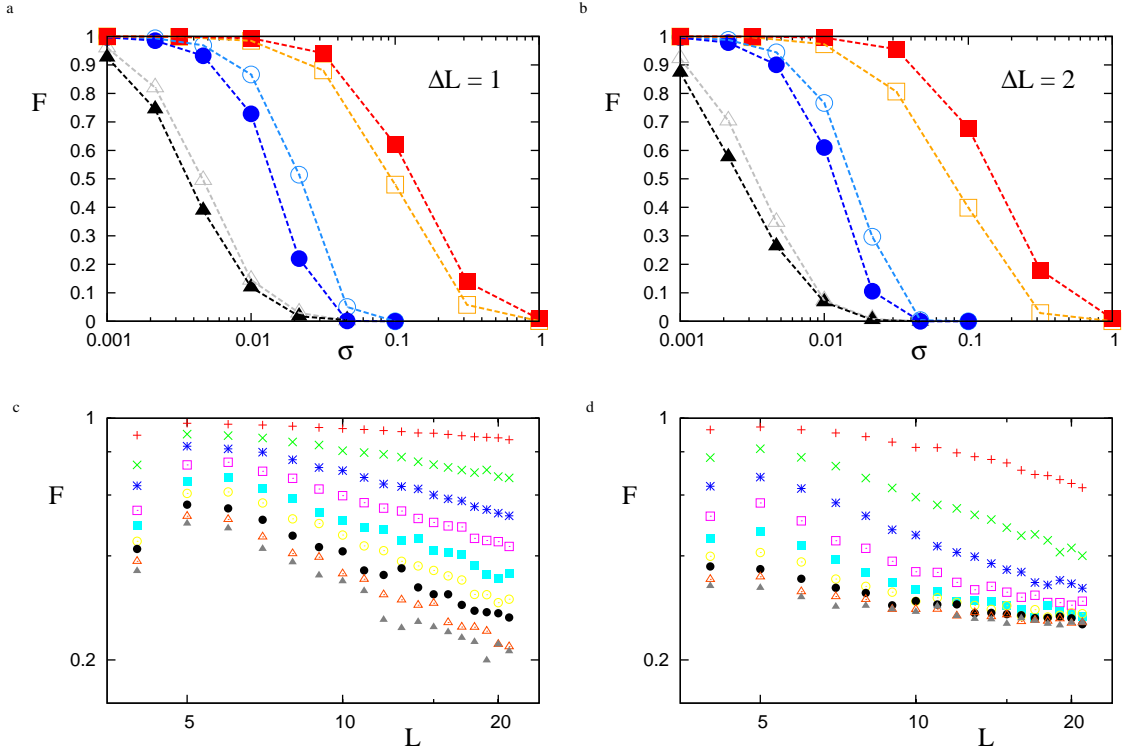}
\addspace
\caption {\label{fig5}
Fidelity $F$ of quantum computers running Shor's algorithm
(a and b), and a modulo addition gate (c and d).
In the order of a and b, the quantum computer is equipped
with adders that are capable of being used in Shor-algorithm
factoring of 5- and 6-bit semiprimes.
Compared to Fig.~\ref{fig1}, the boost from symmetrized errors
is more significant when factoring 15, as shown.
Frames~c and d show $F$ as a function of the bit-length $L$
of the maximal semiprime that may be factored using
a modulo-addition gate, equipped with
relative and absolute symmetric errors, respectively.
All cases were performed with $N=15$.
In decreasing order of $F$,
different plot symbols refer to
$\sigma = 0.1 ,0.2, \ldots, 0.9$.
}
\end{figure}

We also note in passing that we observe an extra boost of fidelity
when we introduce more qubits to the quantum circuit than necessary (see Fig.~\ref{fig5}).
We find the smallest subcircuit that exhibits such an extra boost to be
the modulo addition gate, whose fidelity as a function 
of the number of extra qubits
$\Delta L$ appears in Figs.~\ref{fig5} c and d.
In fact, in Fig.~\ref{fig3}, different color symbols represent
different numbers of extra qubits used in the modulo-addition gate,
clearly indicating the presence of this extra boost.

A crude, simple analytical analysis may be performed on the modulo-addition gate
based on our previous results, in order to show this extra boost 
exhibited in Figs.~\ref{fig5} c and d.
To a very crude approximation, the limiting distribution $|R(x)|^2$ in (\ref{Rx}),
in the limit of small $\sigma$, may be approximated as a delta-peak centered at
the ideal output $x_0$ with a uniform distribution throughout 
the rest of the domain
of the integral in (\ref{Premain}), such that $\int_0^1 |R(x)|^2 dx = 1$.
Now, $R$ in (\ref{R}) shows that increasing $L$ while keeping addends the same
does not change $R$ for an ideal output.
Thus, together with $|R|^2 \approx F_{\text{adder}} \delta(x-x_0) + 
(1-F_{\text{adder}})$ for $x \in [0,1)$,
we obtain $P_{\text{remain}} \approx F_{\text{adder}} + (1-F_{\text{adder}})
[0.5+\sin(\pi x_0)/\pi]$, where $x_0 = N/2^{L_{\text{min}}+\Delta L}$.
Despite its crudeness, the fidelity of the corresponding circuit
$F_{\text{s.s.}} = P_{\text{remain}}^2$ shows a clear extra-boost behavior
as a function of $\Delta L$, demonstrating the power of our analytical model.

\section*{Discussion}
Clearly, our analytical results scale in the number of qubits,
demonstrating that the symmetry-driven fidelity boost will persist
as we scale up the quantum circuit.
We also notice that the analytically predicted fidelity
[see e.g. (\ref{Ratio}) and (\ref{Fshor})] 
underestimates the numerically observed fidelity boost.
This is so, because our analytical analyses are based on
local estimates of fidelity boosts that are focused on
individual building blocks, such as adders and modulo adders.
Thus, since the boosts in the individual building blocks are
undeniably present, we take the boosts obtained on the basis
of the individual building blocks as a lower limit of the globally
achievable boost, which, as we demonstrated explicitly with 
our Shor algorithm simulations, 
may be as large as a factor 10. 
We expect additional boosts due to long-range coherences
that are not currently contained in our local analytical 
estimates. These need to be investigated 
further in order to identify their origins and working principles.

We are certain that our results are a welcome boon for 
quantum experimentalists and engineers.
Not only is the quantum computer already resilient against irremovable
hardware errors, but, as we showed in this paper,
exhibits significant performance enhancement 
just by controlling the symmetry of the errors. 
We also showed that using symmetry as a method 
to boost performance 
is well within engineering capabilities. 
This is supported by the fact that spin-echoes \cite{Spin-echo},
e.g., already proved useful for practical applications 
in suppressing the naturally occurring errors in a given physical system.
While it is still true that the symmetry needs to be implemented to a high precision,
from the engineering perspective, the task of 
keeping the symmetry should be
easier than keeping the error level itself small.
Our results are also of interest to theorists.
Given that exploiting symmetry is the key 
for the dramatic fidelity boost
at the architectural, surface level, 
as opposed to the individual, microscopic,
inner-workings of a single-qubit state,
we gain the insight that 
a topologically and structurally robust 
quantum algorithm may be developed.
Given the fact that quantum algorithms, in general,
tend to contain a large number of symmetric structures,
we expect that designing hardware that 
results in symmetric errors, as exploited 
in this paper, may be beneficial for boosting 
performance in other quantum
algorithms as well. 

It would have been lamentable if the irremovable 
hardware errors in the logical qubits proliferated too quickly
for a quantum computer to be of practical use.
Fortunately, as we showed in this paper, this is not so.
Together with the pioneering works in quantum error correction
and its fault-tolerant implementation,
the surprising robustness of quantum computers
with respect to errors and noise suggests
that quantum computing and quantum information
are more than just of academic interest. Exploiting 
symmetry in the subunits of quantum algorithms 
as suggested in this paper provides an additional   
powerful tool on the way to the construction of 
quantum computers of practical importance. 

\section*{ADDITIONAL INFORMATION}
{\bf Competing financial interests:} The authors 
declare no competing financial interests.


\section*{AUTHOR CONTRIBUTIONS}
Y. S. Nam and R. Bl\"{u}mel devised the idea and
wrote the main manuscript.
Y. S. Nam analyzed the results.

\end{document}